\documentstyle[emulateapj]{article}

\newcommand{\etal}{{\it et al.~}}
\newcommand{\eg}{{\it e.g.~}}
\newcommand{\Om}{{$\Omega_m$}}
\newcommand{\arcm}{\hbox{$^\prime$}}
\newcommand{\arcs}{\hbox{$^{\prime\prime}$}}     

\begin{document}

\title{Discovery of a Galaxy Cluster via Weak Lensing}

\author{D. Wittman, J.~A. Tyson, V. E. Margoniner} 
\affil{Bell Laboratories, Lucent Technologies, Murray Hill,
NJ 07974 \\ wittman,tyson,vem@physics.bell-labs.com}
\author{J.~G. Cohen}
\affil{Astronomy Department, California Institute of Technology,
Pasadena, CA 91125 \\ jlc@astro.caltech.edu }
\author{I.~P. Dell'Antonio} 
\affil{Physics Department, Brown University, Providence, RI 02912 \\
ian@het.brown.edu}

\begin{abstract}
We report the discovery of a cluster of galaxies via its weak
gravitational lensing effect on background galaxies, the first
spectroscopically confirmed cluster to be discovered through its
gravitational effects rather than by its electromagnetic radiation.
This fundamentally different selection mechanism promises to yield
mass-selected, rather than baryon or photon-selected, samples of these
important cosmological probes.  We have confirmed this cluster with
spectroscopic redshifts of fifteen members at z=0.276, with a velocity
dispersion of 615 km s$^{-1}$.  We use the tangential shear as a
function of source photometric redshift to estimate the lens redshift
independently and find $z_l = 0.30 \pm 0.08$.  The good agreement with
the spectroscopy indicates that the redshift evolution of the mass
function may be measurable from the imaging data alone in
shear-selected surveys.
\end{abstract}


\section{Introduction}

Clusters of galaxies are essential tools for developing our
understanding of structure formation and for probing cosmological
parameters.  In particular, the redshift evolution of the cluster mass
function is a sensitive diagnostic of \Om, sufficiently sensitive that
the existence of even one or two massive clusters at $z \sim 0.8$
favors a low-density universe (Donahue \etal 1998, Ebeling \etal
2000).  This argument assumes Gaussianity in the primordial
fluctuations; clusters are equally useful at constraining primordial
non-Gaussianity given an independent measure of \Om\ (Robinson,
Gawiser \& Silk 1999).  Precise estimates of either quantity will
require large, unbiased samples of clusters at a range of redshifts.
Locally, unbiased samples are crucial for measuring $\sigma_8$
and for estimates of \Om\ that rely on the fair sample hypothesis,
i.e. that the composition of clusters in terms of baryon fraction or
mass-to-light ratio is representative of the rest of the universe
(Evrard 1997, Carlberg \etal 1996).  Finally, the evolution of the
numbers of cluster-sized masses as a function of redshift constrains
the cosmological constant and the dark energy equation of state
(Tegmark 2001).

If clusters are to serve as measures of the clumping of mass, we must
identify them directly from observations of the mass distribution.
This is difficult, however, because the vast majority of their matter
is dark.  The traditional methods of discovering clusters rely on
optical emission from galaxies (\eg Abell, Corwin \&
Olowin 1989; Zaritsky, \etal 1997) or X-ray emission from a hot
intracluster plasma (\eg Boehringer \etal 2000).  Clusters
with a lower mass-to-light ratio or more baryons may well be
overrepresented in these samples.  Radiation from baryons is a
complicated proxy for mass, and hydrodynamic baryon-CDM models
have been proposed to study bias in clusters (Blanton \etal 1999).

The development of weak gravitational lensing techniques (see Mellier
1999 and Bartelmann \& Schneider 2001 for reviews) suggests a
different approach: shear selection (Schneider 1996).  Because all
types of matter participate in lensing, deep, wide-field imaging will
reveal in its shear pattern all mass concentrations regardless of
mass-to-light ratio or baryon fraction. Of course, no single technique
will be completely unbiased. For example, shear selection, like
optical selection and the Sunyaev-Zel'dovich effect (Joy \& Carlstrom
2001), will tend to be more sensitive to line-of-sight superpositions
of unrelated structures, to which X-ray selection, with its dependence
on the square of the density, is less vulnerable.  With X-ray and
(until now) shear selection, optical followup is still necessary to
provide a redshift.  Comparison of differently selected samples will
always be necessary.  Still, the baryon and photon independence
provided by shear selection are powerful features which may
produce surprising new samples.

Two ``dark'' mass concentrations, found through weak lensing analyses
of apparently unrelated previously known clusters, have been reported
(Erben \etal 2000; Umetsu \& Futamase 2000).  Due to the absence of
associated galaxies, the redshift and therefore masses of those clumps
remain unknown.  We report here the first spectroscopically confirmed
shear-selected cluster.  We also introduce photometric redshift
techniques into the selection of sources for the weak lensing
analysis.  The recent development of these techniques (see Connolly
\etal 1995; Hogg \etal 1998) greatly improves the promise of shear
selection, as source galaxies may be divided into redshift bins to
tune the sensitivity to lensing mass concentrations through a series
of lens redshifts (mass tomography).  In this paper, we use
photometric redshifts to demonstrate that the shear-selected mass
concentration is at roughly the same redshift as the cluster galaxies.
This has always been an assumption in lensing analyses of clusters,
but this measurement implies that mass tomography is feasible.

\section{Imaging Observations and Lensing Detection}

We took $B_jVRI$ images of a ``blank'' field (containing no known
cluster) centered at 23:47:46 +00:57:42 (J2000) using the Big
Throughput Camera (BTC, Wittman \etal 1998) in photometric conditions
at the Blanco 4-m telescope of Cerro Tololo Inter-American Observatory
in 1997 and 1998.  Details of the data reduction, galaxy catalogs and
seeing correction may be found in Wittman \etal (2000, W00); here we
give only the basic parameters.  The final images are roughly 40\arcm\
square with 0.43\arcs\ sampling and 1.3\arcs\ FWHM after point-spread
function corrections, with object counts peaking at $Bj=26.4$,
$V=26.1$, $R=25.6$, and $I=24.4$ (isophotal magnitudes).  In the final
catalog, each object's shape is a weighted mean of the shapes measured
in the different filters, as described in W00.

In addition to the W00 magnitude cut of $23 < R < 26$, we imposed a
color cut of $B_j - R < 1.5$ to emphasize blue field galaxies at
higher redshift, a tactic often used in weak lensing analyses of known
clusters (Tyson, Wenk \& Valdes 1990).  Using the method of Fischer \&
Tyson (1997), we constructed a mass map from the remaining 31,000
galaxies (Figure 1).  A mass concentration stands out near the
southwest corner; its peak is significant at the 4.5$\sigma$ level,
based on mass maps of bootstrap resampled catalogs.  The mass
concentration disappeared, as it should, under two null tests:
rotating each galaxy in the catalog by 45$^\circ$, and assigning the
shape of a random different galaxy to each galaxy position.  To check
for any bias casued by proximity to the field edge, we simulated a
field of the same size and plate scale filled with random galaxies,
distorted the image as if lensed by a cluster of moderate ($800$ km
s$^{-1}$) velocity dispersion, convolved and added noise to match the
seeing and noise of the observations, cataloged, selected sources, and
made mass maps as for the observations.  Repeating this for a series
of ten center-to-corner cluster positions, we found that the bias was
small ($<10\%$ in peak density) and in the sense of reducing, not
enhancing, our sensitivity to mass concentrations near the corner.

We then made a color composite image from the $B_j$ and $R$ images
(Figure 1.5, separate jpeg).  A concentration of reddish elliptical
galaxies appears near the position of the density peak (23:46:23.85
+00:45:00.8 for the brightest cluster galaxy versus 23:46:24.0
+00:43:58.8 for the density peak, a displacement of just over 1\arcm).
Nothing appears at that position in the ROSAT database, and the
NASA/IPAC Extragalactic Database contains only one object in the area,
a large spiral which appears to be a foreground field galaxy.  We
concluded that this candidate cluster was worthy of spectroscopic
followup.

\section{Spectroscopic Followup}
\label{sec-spec}
We designed a single slitmask containing 26 objects believed to be
cluster members, selected by avoiding blue objects and choosing
suitably bright yellow/red galaxies in the color composite image in
the area around the putative cluster.  This slitmask was used with the
Low Resolution Imaging Spectrograph (Oke \etal 1995) on the Keck
Telescope on the night of Nov. 23, 2000.  A single 1200 sec exposure
with the 300 l/mm grating was obtained; the spectral resolution was 10
\AA\ with a 1 arc-sec wide slit, and the region from 4000 to 8000 \AA\
was covered for each object.  Two of the objects proved to be Galactic
M dwarfs, while 17 are members or near outliers of a cluster at
z=0.276.  One is a foreground galaxy, and six are background galaxies.
We assume an instrumental contribution to the velocity dispersion of
100 km/sec in the rest frame.  While there are two outliers among the
17 possible cluster members, both sigma-clipping and the biweight
estimator of Beers, Flynn \& Gebhardt (1990) yield a velocity
dispersion estimate (in the rest frame) of $\sigma_v = 615 \pm 150$ km
s$^{-1}$.  

\section{Lens Redshift and Mass Estimates Using Photometric Redshifts}
\label{sec-lensmass}

With the cluster redshift in hand, we need only the redshift
distribution of the lensed sources to derive the cluster mass from the
shear.  In the past it has been difficult enough to estimate the mean
of this distribution, but photometric redshifts can provide a detailed
distribution, in principle even an appropriate weight for each source
galaxy.  In this paper, we use tangential shear as a function of source
photometric redshift to estimate the lens redshift $z_l$ in a way
independent of the spectroscopy.  We demonstrate that the mass causing
the shear signal is at roughly the same redshift as the cluster.

For each galaxy observed in all four filters, we used the HyperZ
package (Bolzonella, Miralles \& Pell\'{o} 2000) to compute a redshift
probability density function (PDF).  We multiplied this by another PDF
computed from the galaxy's apparent magnitude, assuming that a
Schechter (1976) luminosity function with $M_{B_j}^* = -19.73$ and
$\alpha = -1.28$ (Folkes \etal 1999) holds at each redshift (we assume
$H_0 = 70$ km s$^{-1}$ Mpc$^{-1}$, $\Omega_m = 0.3$, and
$\Omega_{\Lambda} = 0.7$ throughout).  The latter PDF is quite
broad, but serves a valuable purpose by suppressing high-redshift
peaks which often appear in the former PDF.  The first and second
moments of the final PDF product provide the estimated redshift and
its statistical error.  Henceforth, we use a catalog of 26,000
galaxies which is defined not by the magnitude and color cuts that
went into Figure 1, but by the requirement of detection in each of
four filters leading to a photometric redshift.  Of these 26,000
galaxies, 13,000 were also in the initial catalog.  The median $z_{\rm
phot}$ in this catalog is 0.58.

We verified the accuracy of the photometric redshifts by comparison
with spectroscopic redshifts of 31 galaxies in the range $ 0.23 <
z_{\rm spec} < 0.83$.  This sample was composed of 19 galaxies with
redshifts described in Section~\ref{sec-spec} (the ones detected in
all four filters), plus another 12
galaxies ($ 0.24 < z < 0.83$) in a different region of the 40\arcm\
field, kindly provided by R. Guhathakurta.  We find little bias, with
$(z_{\rm spec} - z_{\rm phot}) / (1 + z_{\rm spec}) = -0.027$ on
average, and an rms of only 0.059 in the same quantity.  A
detailed analysis of this photometric redshift method and its
application to larger datasets is in preparation (Margoniner \etal
2001).

Before examining the dependence of the shear on source redshift, we
must first account for the equally large dependence on projected
position relative to the lens.  We are not interested in a detailed
reconstruction of the lens; rather, for a given $z_{\rm phot}$ we
would like to collapse information from all sources at a wide range of
projected radii into a single number characterizing the lensing signal
at that source redshift.  Hence at a given $z_{\rm phot}$ we fit a
very simple model, a singular isothermal sphere (SIS), to the radial
dependence of the tangential shear (centered on the peak in the mass
map), and use the fitted amplitude and its uncertainty.  The
assumption of a particular profile should not introduce any bias as a
function of $z_{\rm phot}$.  To test this, we also considered NFW
(Navarro, Frenk \& White 1997) profiles.  Although our data cannot
constrain the scale radius $r_s$, for fixed $r_s$ (225 kpc,
Bartelmann, King \& Schneider 2001), the results do not change
significantly from the SIS case.  We therefore choose the SIS for
simplicity.  For an SIS, the tangential shear $\gamma_t(r) = \Sigma(r) /
\Sigma_{\rm crit}$, where $\Sigma(r) \sim r^{-1}$ is the projected
mass density, $\Sigma_{\rm crit} = (c^2 / 4\pi G) (D_s / D_{ls} D_l)$
is the critical density, and $D_s$, $D_{ls}$, and $D_l$ are the
angular diameter distances from observer to source, lens to source,
and observer to lens, respectively.  Since only $D_s$ and $D_{ls}$ are
changing with source redshift, the amplitude of an $r^{-1}$ fit to the
tangential shear should grow with the distance ratio $D_s / D_{ls}$.

Figure 2 shows the fitted tangential shear (at a fiducial radius of 1
Mpc) as a function of $z_{\rm phot}$.  It is consistent with
zero for $z_{\rm phot} \le 0.3$ and increases monotonically for $ 0.3
< z_{\rm phot} < 1.3$ (the upper limit is to avoid extrapolating too
far beyond the range of spectroscopic verification of $z_{\rm phot}$).
The dotted line illustrates the shear expected from a lens at $z_l =
0.276$, with $\Sigma$ fit to the points.  This is a good fit
($\chi_\nu^2 \sim 1$).  We explore the range of $z_l$ allowed by the
shear data by stepping $z_l$ through the range $0.025 \le z_l \le 1.3$
in steps of 0.025, and refitting at each step.  The probabilities
corresponding to $\chi^2$ at each step are plotted in Figure 3.  The
median and mode of this distribution are at $z_l = 0.31$ and 0.30
respectively, with a 68\% confidence interval $0.225 < z_l < 0.375$
(these numbers change by less than 0.01 when an NFW profile is used).
The fit for $z_l = 0.3$ is also shown in Figure 2 (dashed line).  Thus
the lens roughly coincides with the cluster of galaxies in redshift as
well as in right ascension and declination.  This method can be used
to estimate the redshift of any newly discovered lensing mass from the
lensing data alone.  Such a procedure may well become a standard part
of shear-selected cluster surveys.

Finally, we estimate the mass and mass-to-light (M/L) ratio using the
best-fit projected mass.  Still assuming an isothermal sphere, the
projected mass inside of 250 kpc (where it is convenient to measure
the light) is $2.8 \pm 0.6 \times 10^{14} M_\odot$, assuming $z_l =
0.276$.  For the range of $z_l$ allowed by the $\gamma_t(z_{\rm phot})$
curve, $M_{\rm proj}(<250\ {\rm kpc})$ ranges from $1.8-3.7 \times
10^{14} M_\odot$.  The velocity dispersion implies $M_{\rm proj}(<250\
{\rm kpc}) = 0.7 \pm 0.3 \times 10^{14} M_\odot$ under the SIS
assumption.  The discrepancy may be due to the SIS assumption: Unlike
the redshift estimate, the mass estimate is sensitive to the profile
assumed.  Converting from observed $I$ band
to rest-frame $R$ using the approach of Fischer \& Tyson (1997), we
find $M/L_R = (560 \pm 200) h$ in this region.  Compared to other
clusters (Mellier 1999), this is high but not exceptional.

\section{Discussion}

Baryon-unbiased samples of mass concentrations over a wide range of
redshift will be of critical importance in constraining cosmological
parameters.  To realize the potential of this technique, weak lensing
observations must have the sensitivity to discover clusters over a
broad part of the cluster mass function.  For example, constraints on
\Om\ from the mass function at high redshift currently involve only a
few very massive clusters, and such extreme clusters lie far out on
the tail of the mass distribution, which may not be Gaussian.
Weak lensing surveys can probe clusters an order of magnitude less
massive.  Unlike X-ray and optical selection, a shear signal does not
diminish as the square of the luminosity distance, so that low-mass
clusters should be detectable even at high redshift as long as the 
photometric redshifts are accurate in eliminating foreground
sources.

We have demonstrated the serendipitous discovery, with high
signal-to-noise, of a rather modest cluster via a weak lensing
analysis of a single 40\arcm\ field.  The cluster was
spectroscopically confirmed at $z = 0.276$, with a velocity dispersion
of 615 km sec$^{-1}$.  The tangential shear follows source
(photometric) redshift in a manner which requires the lens to lie at
or near the cluster redshift.  Thus, all the ingredients are in place
for a truly shear-selected sample of clusters, in which any putative
mass clump can be confirmed, and its redshift estimated, from the
multi-color imaging data alone.  This technique is also capable of
answering the question of the existence of any truly dark clumps.  The
redshift (and therefore mass, and mass-to-light ratio lower limit) of
any such clumps can only be derived from the shear versus source
redshift curve.

This further suggests that the promise of three-dimensional mass
tomography (Tyson 1995, 2000) over cosmologically significant volumes
can be realized in wide-field deep imaging surveys.  Note that such a
cluster is not unexpected in the volume probed by this data
(Rahman \& Shandarin 2001).
Ongoing cosmic shear surveys covering tens of square degrees ({\it
e.g.} the Deep Lens Survey\footnote{http://dls.bell-labs.com},
DESCART\footnote{http://terapix.iap.fr/Descart}) should discover
significant samples of shear-selected clusters (Kruse \& Schneider
1999) and begin to constrain
\Om\ and dark energy through the redshift evolution of the cluster
mass function.

\acknowledgments

We thank P. Guhathakurta for providing spectroscopic redshifts outside
the cluster, and the CTIO staff for their help with the BTC project
and for their upgrading and maintenance of the delivered image quality
of the Blanco telescope.  Cerro Tololo Inter-American Observatory is a
division of National Optical Astronomy Observatory (NOAO), which is
operated by the Association of Universities for Research in Astronomy,
Inc., under cooperative agreement with the National Science
Foundation.  BTC construction was partially funded by the NSF.

\clearpage

\begin{figure}
\plotone{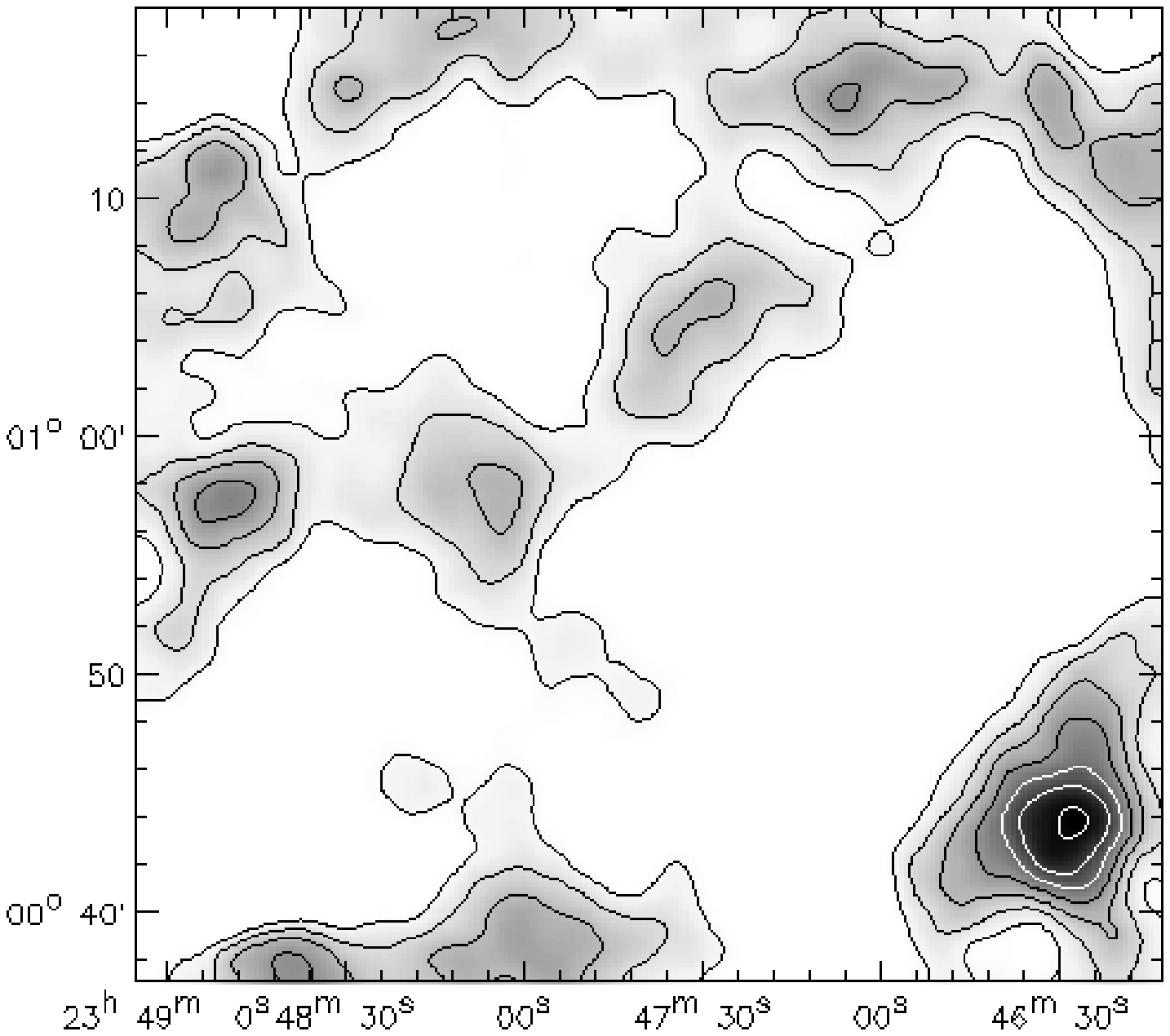}
\caption{Projected mass map, smoothed on a 2\arcm\ scale, of the
40\arcm\ field.  Higher-density regions are shown darker.  Contours
are equally spaced in arbitary units (but linear in projected mass
density); negative contours are omitted for clarity.  One peak, at
lower right, stands out at twice the density of any other peak
(4.5$\sigma$).  This mass overdensity corresponds to a small cluster
of galaxies spectroscopically confirmed at z=0.276.  The width of this
field at that redshift is 13 Mpc.  North is up and east is to the
left.  }
\end{figure}

\begin{figure}
\plotone{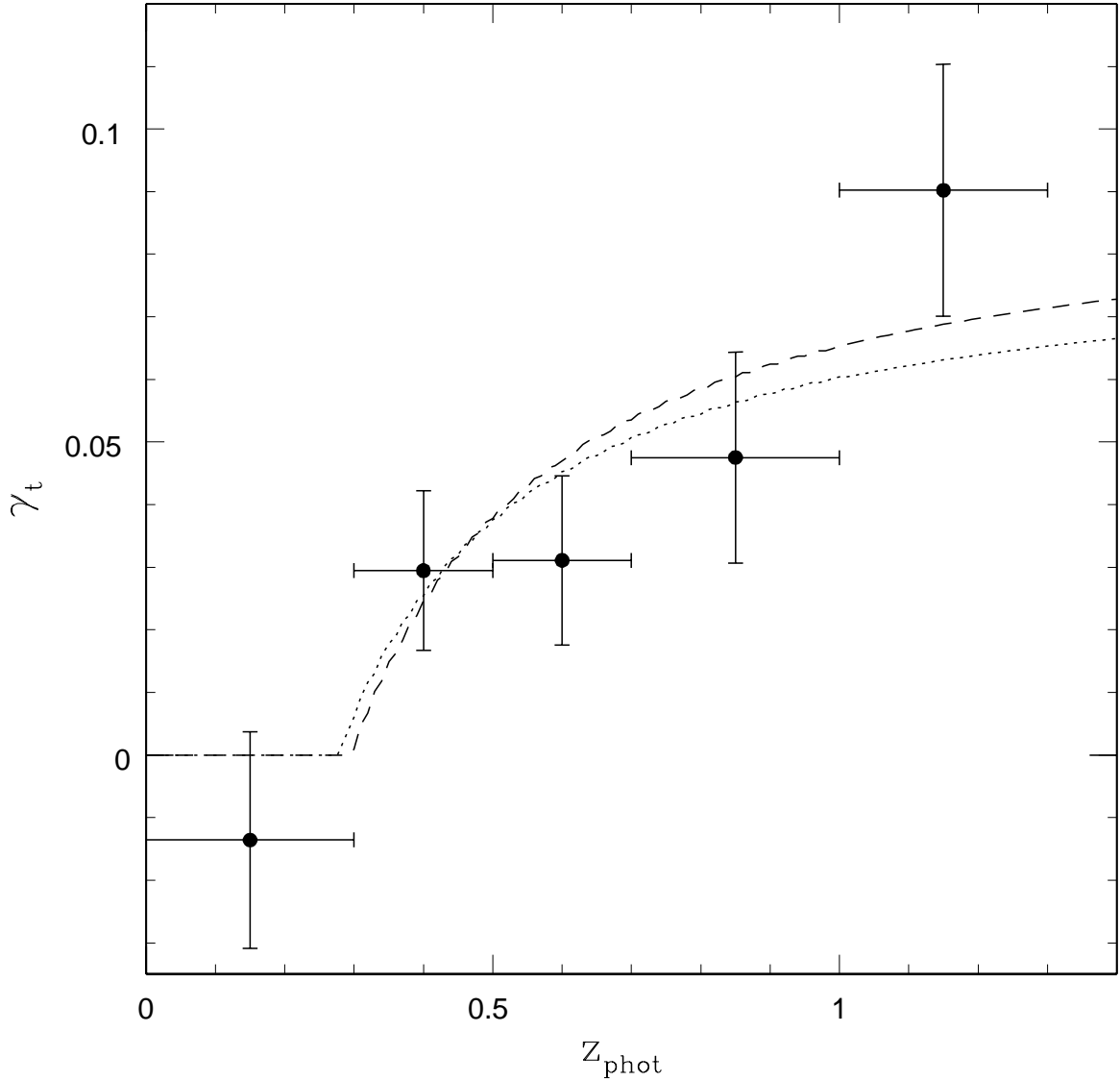}
\caption{Measured shear as a function of source photometric redshift
(points).  Each point represents the amplitude of a best-fit
isothermal shear profile at a fiducial radius of 1 Mpc, with vertical error
bars indicating the uncertainty in the fit.  The dotted and dashed
lines represent the shear expected from lenses at $z=0.276$ and
$z=0.30$, the spectroscopic and shear-derived redshifts respectively
(the different amplitudes reflect slightly different best-fit masses).
The horizontal error bars represent the nominal widths of the bins
only; the effect of scatter in the photometric redshifts is
neglected. }
\end{figure}

\begin{figure}
\plotone{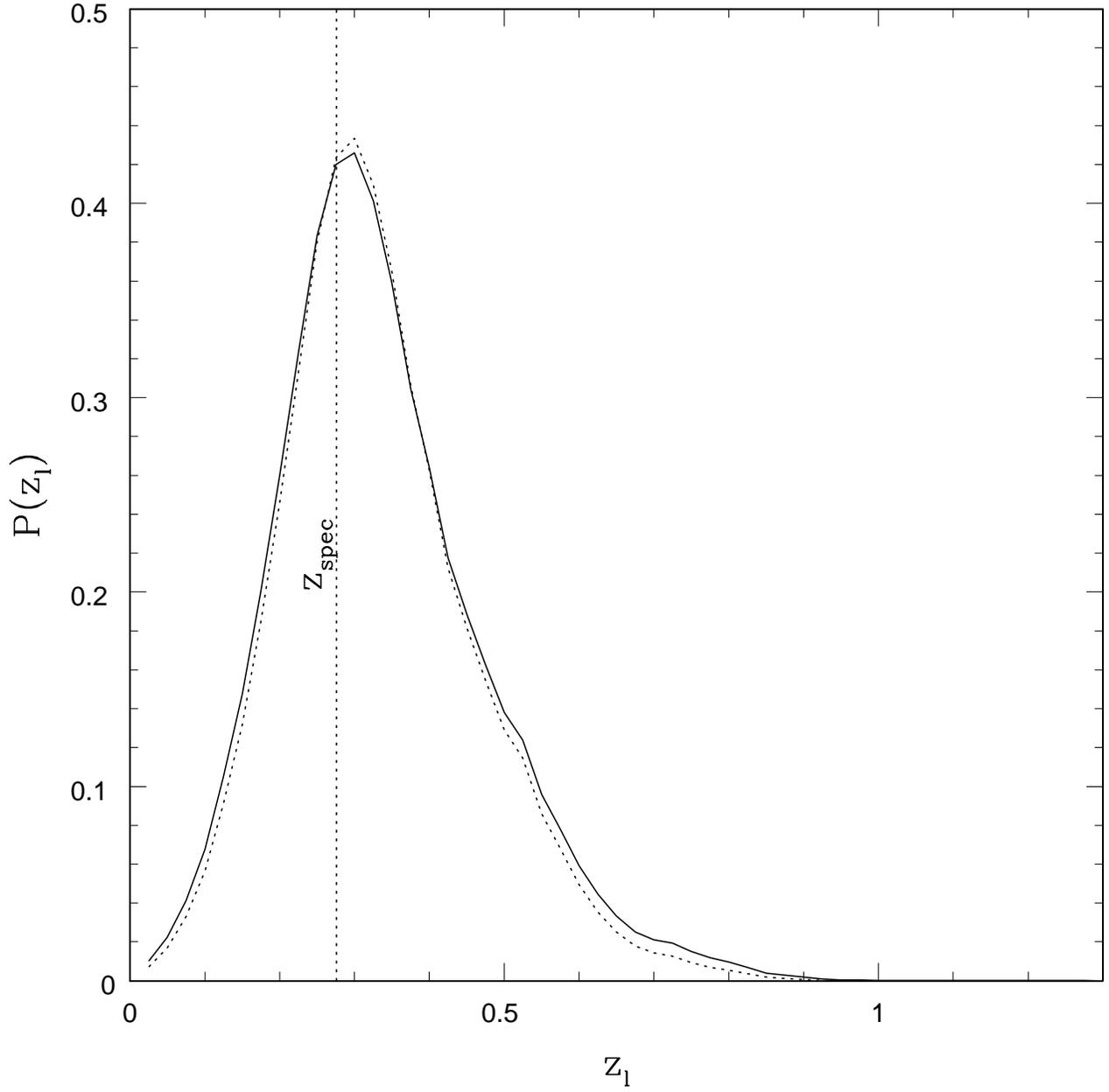}
\caption{Lens redshift probability density function (unnormalized).
Solid line: at each putative lens redshift $z_l$, we fit a lens mass
to $\gamma_t(z_{\rm phot})$ (Figure 2) and compute the probability
from the $\chi^2$ for the remaining four degrees of freedom (five data
points minus one fit parameter).  Dotted line: we repeated the process
assuming an NFW profile with $r_s = 225$ kpc.  Either assumption leads
to a most probable $z_l$ within 0.03 of the spectroscopic value.}

\end{figure}

\end{document}